\documentstyle[aps,preprint,epsf]{revtex}
\newcommand{\beq}{\begin{equation}}
\newcommand{\eeq}{\end{equation}}
\begin{document}
\draft \tightenlines

\title{ Possible Universal Cause of
High-T$_c$ Superconductivity in Different Metals}

\author{M.Ya. Amusia$^{a,b}$, and
V.R. Shaginyan$^{c}$ \footnote{E--mail: vrshag@thd.pnpi.spb.ru}}

\address{$^{a\,}$The
Racah Institute of Physics, the Hebrew University, Jerusalem
91904, Israel;\\ $^{b\,}$ A.F. Ioffe Physical-Technical Institute,
Russian Academy of Sciences, 194021 St. Petersburg, Russia;\\
$^{c\,}$ Petersburg Nuclear Physics Institute, Russian Academy of
Sciences, Gatchina, 188300, Russia} \maketitle

\begin{abstract}
Using the theory of the high temperature superconductivity based
on the idea of the fermion condensation quantum phase transition
(FCQPT), we show that neither the $d$-wave pairing symmetry, nor
the pseudogap phenomenon, nor the presence of the Cu-O$_{2}$
planes are of decisive importance for the existence of the
high-$T_c$ superconductivity. We analyze recent experimental data
on this type of superconductivity in different materials and show
that these facts can be understood within the theory of
superconductivity based on FCQPT. The latter can be considered as
a universal cause of the high-$T_c$ superconductivity. The main
features of a room-temperature superconductor are discussed.
\end{abstract}\bigskip

\pacs{PACS: 71.27.+a; 74.20.Fg; 74.72.-h}

It has been a long-standing problem to develop a theory describing
the high temperature superconductivity (HTS) observed in copper
oxide based compounds. These compounds are extremely complex
materials having a great number of competing degrees of freedom,
which produce a great variety of physical properties. In turn,
these properties can compete and coexist with the
superconductivity hiding the understanding of the universal cause
of the superconductivity. As a result, it was suggested that the
unique superconducting properties in these compounds are
determined by the presence of the Cu-O$_{2}$ planes, by the
$d$-wave pairing symmetry, and by the existence of the pseudogap
phenomena in optimal doped and underdoped cuprates (see e.g.
\cite{tim,tk,vn}).

However, recent studies of quasiparticle tunneling spectra of
cuprates have revealed that the pairing symmetry may change from
$d$ to $s$-wave symmetry, depending on the hole, or electron,
doping level \cite{skin}. Then the high temperature $s$-wave
superconductivity has been observed in electron doped infinite
layer cuprates with a sharp transition at $T=43$ K, without a
pseudogap \cite{chen}. Therefore, we can conclude that the
$d$-wave symmetry of pairs and the pseudogap phenomena are not
necessary parts of HTS. Recent studies of HTS in chemically doped
A$_3$C$_{60}$ \cite{gun} have shown that the presence of the
Cu-O$_{2}$ planes is also not a necessary condition for the
existence of HTS. Moreover, it was recently demonstrated that
strongly overdoped cuprates Tl-2201 obey the Wiedemann-Franz law
perfectly \cite{cyr}. This fact, manifesting that excitations
which carry heat also carry charge $e$, shows that these
quasiparticles have the same properties as Landau quasiparticles.
This imposes serious constrains upon the possible theories of HTS.
For instance, this fact demonstrates that there is no spin-charge
separation \cite{cyr} as was suggested in one group of proposals
\cite{and,kiv}. On the other hand, it is well known that the
superconducting transition temperature $T_c^{\alpha\gamma}(x)$ of
the oxides appears to pass through a bell-shaped curve as a
function of the hole (or electron)
mobile charge carrier density $x$ (see e.g.
\cite{vn}). Here $\alpha$ denotes the material, say C$_{60}$ or
Tl-2201, etc., and $\gamma$ denotes hole or electron doping.  It
is essential to note that the shape of the functions
$T_c^{\alpha\gamma}(x)$ is similar in different samples, so that
there exist empirical formulas, which connect the transition
temperature $T_c^{\alpha\gamma}(x)$ with the carrier concentration
$x$. For example, in case of Tl-2201, see e.g. \cite{cyr},
$T_c(x)=T^M_c(1-82.6(p-0.16)^2)$. Here $T^M_c$ is
the maximum value of the transition temperature.  Thus, we can use
a simple approximation \beq T_c^{\alpha\gamma}(x)
=T_1^{\alpha}T_2^{\gamma}(x_1-x)x,\eeq where the coefficients
$T_1^{\alpha}$ and $T_2^{\gamma}$ define the transition
temperature $T_c$ for a given hole (or electron) metal,
with $x$ obviously going continuously to
zero at the insulator-metal transition. It directly follows from
Eq. (1) that the transition temperature reaches its maximum value
$T^M_c$ at the optimal doping level $x_{opt}$
\beq T^M_c=T_c^{\alpha\gamma}(x_{opt})=T_1^{\alpha}T_2^{\gamma}
\left(\frac{x_1}{2}\right)^2.\eeq The general shape of the
function $T_c(x)$ points to the fact that the generic properties
of HTS are defined by the 2D charge (electron or hole) strongly
correlated liquid rather then by solids which hold this liquid. As
to the solids, they arrange the presence of the pseudogap
phenomena, $s$ or $d$-wave pairing symmetry, the electron-phonon
coupling constant defining $T_c$, the variation region of $x$, and
so on.

It is well-known that consideration of strongly correlated liquids
is close to the problem of systems with a big coupling constant,
which is the most important in many-body physics. One solution of
this problem has been offered by Landau theory of normal Fermi
liquids, which introduce the notion of quasiparticles and
parameters characterizing the effective interaction between them
\cite{lan}.  As a result, the Landau theory has removed high
energy degrees of freedom and kept a sufficiently large number of
relevant low energy degrees of freedom to treat the liquid's low
energy properties. Usually, it is assumed that the breakdown of
the Landau theory is defined by the Pomeranchuk stability
conditions and occurs when the Landau amplitudes being negative
reach critical value. The new phase, in which the stability
conditions are restored can in principle be again described in the
framework of the theory. However, it was demonstrated recently
\cite{ks} that the Pomeranchuk's conditions do not cover all
possible limitations: the overlooked one is connected to the
situation when, at temperature $T=0$, the effective mass can
become infinitely big. It has been demonstrated that such a
situation, leading to profound consequences, can take place when
the corresponding amplitude, being positive, reaches a critical
value, producing a completely new class of strongly correlated
Fermi liquids with the fermion condensate (FC) \cite{ks,vol}. This
liquid is separated from a normal Fermi liquid by the fermion
condensation quantum phase transition (FCQPT) \cite{ms}. In such a
case we are dealing with the strong coupling limit where an
absolutely reliable answer, based on pure theoretical first
principle ground, cannot be given. Therefore, the only way to
verify whether the FC occurs is to consider experimental facts,
which can signal to the existence of such a state. It seems to us
that there are strong indications on FC existence in high-$T_c$
superconductors.

In this Letter we show that the above mentioned new experimental
facts are the desirable evidences and that they can be explained
within the theory of HTS based on FCQPT \cite{ams}. We also
describe the main features of corresponding room-temperature
superconductors.

Let us start with a brief consideration of the general properties
of a two-dimensional electron liquid in the superconducting state,
when the system has undergone FCQPT \cite{ms,ams}. At $T=0$, the
ground state energy $E_{gs}[\kappa({\bf p}),n({\bf p})]$ is a
functional of the order parameter of the superconducting state
$\kappa({\bf p})$ and of the quasiparticle occupation numbers
$n({\bf p})$ and is determined by the well-known equation of the
weak-coupling theory of superconductivity (see e.g. \cite{til})
\beq E_{gs}=E[n({\bf p})] +\int \lambda_0V({\bf p}_1,{\bf
p}_2)\kappa({\bf p}_1) \kappa^*({\bf p}_2) \frac{d{\bf p}_1d{\bf
p}_2}{(2\pi)^4}.\eeq Here  $E[n({\bf p})]$ is the ground-state
energy of a normal Fermi liquid, $n({\bf p})=v^2({\bf p})$ and
$\kappa({\bf p})=v({\bf p})\sqrt{1-v^2({\bf p})}$. It is assumed
that the pairing interaction $\lambda_0V({\bf p}_1,{\bf p}_2)$ is
weak. By minimizing $E_{gs}$ with respect to $\kappa({\bf p})$ we
obtain the equation connecting the single-particle energy
$\varepsilon({\bf p})$ to $\Delta({\bf p})$ \beq \varepsilon({\bf
p})-\mu=\Delta({\bf p}) \frac{1-2v^2({\bf p})} {2\kappa({\bf
p})}.\eeq The single-particle energy $\varepsilon({\bf p})$ is
determined by the Landau equation, $\varepsilon({\bf p})=\delta
E[n({\bf p})]/\delta n({\bf p})$ \cite{lan}, and $\mu$ is the
chemical potential. The equation for the superconducting gap
$\Delta({\bf p})$ takes form \beq \Delta({\bf p})
=-\int\lambda_0V({\bf p},{\bf p}_1)\kappa({\bf p}_1) \frac{d{\bf
p}_1}{4\pi^2}.\eeq If the coupling constant $\lambda_0\to 0$, the
maximum value $\Delta_1$ of the superconducting gap  tends to
zero, $\Delta_1\to 0$, and Eq. (4) reduces to that proposed in
\cite{ks} \beq \varepsilon({\bf p})-\mu=0,\: {\mathrm {if}}\,\,\,
\kappa({\bf p})\neq 0,\,\, (0<n({\bf p})<1); \,\,\: p_i\leq p\leq
p_f\in L_{FC}.\eeq At $T=0$, Eq. (6) defines a new state of a
Fermi liquid with FC, for which the modulus of the order parameter
$|\kappa({\bf p})|$ has finite values in $L_{FC}$ range of momenta
$p_i\leq p\leq p_f$ occupied by FC, and $\Delta_1\to 0$ in
$L_{FC}$ \cite{ks,ms}. Such a state can be considered as
superconducting, with an infinitely small value of $\Delta_1$ so
that the entropy of this state is equal to zero. This state,
created by the quantum phase transition, disappears at $T>0$
\cite{ms}.

It follows from Eq. (6) that the system brakes into two
quasiparticle subsystems: the first one in the $L_{FC}$ range is
occupied by the quasiparticles with the effective mass
$M^*_{FC}\propto 1/\Delta_1$, while the second by quasiparticles
with finite mass $M^*_L$ and momenta $p<p_i$.  It is seen from Eq.
(6) that at the point of FCQPT $p_f\to p_i\to p_F$,  the effective
mass is as large as $1/\Delta_1$.  Equation (6) acquires
nontrivial solutions at $r_s=r_{FC}$ and FCQPT takes place if the
Landau amplitudes depending on the density are positive and
sufficiently large, so that the potential energy prevails over the
kinetic energy and the transformation of the Fermi step function
$n(p)=\theta(p_F-p)$ into the smooth function defined by Eq. (6)
becomes possible \cite{ks}.

Looking now for a simple situation where we can solve Eqs. (4) and
(5) analytically, we assume that $\lambda_0\neq0$ and is small, so
that we can employ the standard theory of superconductivity. In
that case $\Delta_1$ becomes finite, leading to finite value of
the effective mass $M^*_{FC}$ in $L_{FC}$, which can be obtained
from Eq. (4) \cite{ms} \beq M^*_{FC} \simeq
p_F\frac{p_f-p_i}{2\Delta_1},\eeq while the effective mass $M^*_L$
is disturbed weakly.  Here $p_F$ is the Fermi momentum.  It
follows from Eq. (7) that the quasiparticle dispersion can be
presented by two straight lines characterized by the effective
masses $M^*_{FC}$ and $M^*_L$ respectively. These lines intersect
near the binding energy $E_0$ of electrons, which defines an
intrinsic energy scale of the system:  \beq E_0=\varepsilon({\bf
p}_f)-\varepsilon({\bf p}_i)
\simeq\frac{(p_f-p_i)p_F}{M^*_{FC}}\simeq 2\Delta_1.\eeq Equations
(7) and (8) lead to the following general result for the maximum
value $\Delta_1$ coming from the contribution of FC \cite{ams}
\beq\Delta_1\simeq
\beta\varepsilon_F\frac{p_f-p_i}{p_F}\ln(1+\sqrt{2}),\eeq where
$\varepsilon_F=p_F^2/2M^*_L$ is the Fermi energy and $\beta$ is
the dimensionless coupling constant. If the FC contribution to
$\Delta_1$ becomes comparatively small then $\Delta_1$ is given by
the well known relation, being proportional to the exponent of
$(-1/\beta)$, $\Delta_1\propto \exp(-1/\beta)$.

In fact, as it is seen from Eqs. (4) and (5), a
Fermi liquid with FC is
absorbed by the superconducting phase transition and never
exhibits the dispersionless plateau associated with $M^*_{FC}\to
\infty$. As a result, a Fermi liquid
beyond the point of FCQPT can be described by two types of
quasiparticles characterized by two finite effective masses
$M^*_{FC}$ and $M^*_{L}$ respectively, and by the intrinsic energy
scale $E_0$ \cite{ms,ams}. It is reasonably reliable to suggest
that we have returned to the Landau theory by integrating out high
energy degrees of freedom and introducing the quasiparticles. The
sole difference between the Landau Fermi liquid and Fermi liquid
undergone FCQPT is that we have to increase the number of relevant
low energy degrees of freedom by adding both a new type of
quasiparticles with the effective mass $M^*_{FC}$, given by Eq.
(7), and the energy scale $E_0$ given by Eq. (8). We have also to
bear in mind that the properties of these new quasiparticles of a
Fermi liquid with FC cannot be separated from the properties of
the superconducting state, as follows from Eqs. (7), (8) and (9).
We may say that the quasiparticle system in the range $L_{FC}$
becomes very ``soft'' and is to be considered as a strongly
correlated liquid. On the other hand, the system's properties and
dynamics are dominated by a strong collective effect having its
origin in FCQPT and determined by the macroscopic number of
quasiparticles in the range $L_{FC}$.  Such a system cannot be
disturbed by the scattering of individual quasiparticles and has
the features of a quantum protectorate \cite{ms,lpa}. On the other
hand, as soon as the energy scale $E_0\to 0$, the system is driven
back into the normal Landau Fermi liquid \cite{ams}.

Recent studies of photoemission spectra discovered an energy scale
in the spectrum of low-energy electrons in cuprates, which
manifests itself as a kink in the single-particle spectra
\cite{vall,blk}. The spectra in the energy range (-200---0) meV
can be described by two straight lines intersecting at the binding
energy $E_0\sim(50-70)$ meV \cite{blk}. Then, in underdoped copper
oxides, there exists the pseudogap phenomenon and the shape of
single-particle excitations strongly deviates from that of normal
Fermi liquid (see, e.g. \cite{tim}). In the highly overdoped
regime slight deviations from the normal Landau Fermi liquid are
observed \cite{cyr,val1}. All these peculiar properties are
naturally explained within a model proposed in \cite{ms,ams,ms1}
and allow one to relate the doping level, or the charge carriers
density $x$, regarded as the density of holes (or electrons) per
unit area, to the density of Fermi liquid with FC. We assume that
$x_{FC}$ corresponds to the highly overdoped regime at which FCQPT
takes place, and introduce the effective coupling constant
$g_{eff}\sim (x_{FC}-x)/x_{FC}$. According to the model, the
doping level $x$ at $x\leq x_{FC}$ in metals is related to
$(p_f-p_i)$ in the following way:  \beq g_{eff}\sim
\frac{(x_{FC}-x)}{x_{FC}}\sim \frac{p_f-p_i}{p_F} \sim
\frac{p^2_f-p^2_i}{p^2_F}.\eeq

Consider now two possible types of the superconducting gap
$\Delta({\bf p})$ given by Eq. (5) and defined by interaction
$\lambda_0V({\bf p},{\bf p}_1)$. If this interaction is dominated
by a phonon-mediated attraction, the even solution of Eq. (5) with
an $s$-wave, or an $s+d$ mixed waves, will have the lowest energy.
If the pairing interaction $\lambda_0V({\bf p}_1,{\bf p}_2)$ is a
combination of attractive and repulsive interaction, a $d$-wave
odd superconductivity can take place(see e.g. \cite{abr}). But
both an $s$-wave even and $d$-wave odd symmetries lead to
approximately the same values of the gap $\Delta_1$ in Eq. (9)
\cite{ams}. Therefore, the non-universal pairing symmetries in HTS
are likely the results of the pairing interaction, and the
$d$-wave pairing symmetry cannot be considered as absolutely
necessary for HTS existence, in accord with experimental findings
\cite{chen}. If only $d$-wave pairing would exists, the transition
from superconducting gap to pseudogap can take place, so that the
superconductivity will be destroyed at the temperature $T_c$, with
the superconducting gap being smoothly transformed into the
pseudogap, which closes at some temperature $T^*>T_c$
\cite{ms1,sh}.  In the case of $s$-wave pairing we can expect that
the pseudogap phenomena do not exists, in accordance with the
experimental observation (see \cite{chen} and references therein).

We turn now to consideration of the maximum value of the
superconducting gap $\Delta_1$ as a function of the mobile charge
carriers density $x$. Being rewritten in terms of $x$ and $x_{FC}$
related to the variables $p_i$ and $p_f$ by Eq. (10), Eq. (9)
takes the form \beq\Delta_1\propto\beta(x_{FC}-x)x.\eeq Here we
take into account that the Fermi level $\varepsilon_F\propto
p_F^2$, the density $x\propto p_F^2$, and thus,
$\varepsilon_F\propto x$. We can reliably assume that
$T_c\propto\Delta_1$ because the empirically obtained simple bell
shaped curve $T_c(x)$ should reflect only a smooth dependence.
Then, instead of Eq. (11) we have \beq T^{\alpha\gamma}_c
\propto\beta^{\alpha}\beta^{\gamma}(x_{FC}-x)x.\eeq In Eq. (12),
we made the natural change $\beta=\beta^{\alpha}\beta^{\gamma}$
since the coupling constant $\beta$ is fixed by the properties of
metal in question. Now it is seen that Eq. (12) coincides with Eq.
(1) producing the universal optimal doping level
$x_{opt}=x_{FC}/2=x_1/2$ in line with the experimental facts.

Let us make here few remarks. In Ref.\cite{skb1}, HTS was observed
in such fullerides as C$_{60}$, CHBr$_{3}$, and CHCl$_{3}$. In the
case of hole metals, the $T_c^M$ ratios were measured for
C$_{60}$/CHBr$_{3}$-C$_{60}$/CHCl$_{3}$-C$_{60}$, and the same
ratios were also found for the respective electron metals
\cite{skb1}. It follows from Eq. (12) that among these hole doped
fullerides, the $T_c^M$ ratios for
C$_{60}$/CHBr$_{3}$-C$_{60}$/CHCl$_{3}$-C$_{60}$ have to be the
same as in the case of the respective electron doped fullerides
because the factor $\beta^{\gamma}$, related to hole or electron
doping, drops out of the ratios.  We believe that our calculations
present evidences to the correctness of these experimental facts.
Then, it was also shown that a Fermi liquid with FC,
which exhibits strong deviations from the Landau
Fermi liquid behavior, is driven into the Landau Fermi liquid by
applying a magnetic field $B$. This
field-induced Landau Fermi liquid behavior provides
an $A+BT^2$ dependence of the resistivity $\rho$.
A re-entrance into the strongly correlated
regime, when $\rho$ is a linear function of the temperature $T$, is
observed if the magnetic field $B$ decreases to some critical value
$B_{cr}$ \cite{ps}.
We expect that the similar behavior can be observed in case of
strongly overdoped high-temperature compounds at the temperatures
$T\to 0$. If the superconductivity is depressed by the critical
field $B_c$, the $T^2$ behavior of the resistivity  $\rho$ can be
restored at fields $B_{cr}>B_c$. We assume that this behavior
was observed in overdoped Tl-2201 compounds at millikelvin
temperatures \cite{cyr,mac}.

As an example of the implementation of the previous analysis let
us consider the main features of a room-temperature
superconductor. As follows from Eq. (9), $\Delta_1\sim \beta
\varepsilon_F\propto \beta/r_s^2$.  Noting that FCQPT takes place
in three-dimensional systems at $r_s\sim 20$ and in quasi
two-dimensional systems at $r_s\sim 8$ \cite{ksz}, we can expect
that $\Delta_1$ of 3D system comprises 10\% of the corresponding
maximum value of 2D superconducting gap, reaching values as high
as 60 meV for underdoped crystals with $T_c=70$ \cite{mzo}. On the
other hand, it is seen from Eq. (9) that $\Delta_1$ can be even
larger, $\Delta_1\sim 75$ meV, and one can expect $T_c\sim 300$ K
in the case of the $s$ wave pairing as it follows from the simple
relation $2T_c\simeq \Delta_1$. In fact, we can safely take
$\varepsilon_F\sim 300$ meV, $\beta\sim 0.5$ and
$(p_f-p_i)/p_F\sim0.5$. Thus, we can conclude that a possible
room-temperature superconductor has to be a quasi two-dimensional
structure and has to be an $s$-wave superconductor in order to get
rid of the pseudogap phenomena, which tremendously reduces the
transition temperature. The density $x$ of the mobile charge
carriers must satisfy the condition $x\leq x_{FC}$ and be flexible
to reach the optimal doping level. It is worth noting that the
coupling constant $\beta$ is to be sufficiently big because FC
giving rise to the order parameter $\kappa({\bf p})$ does not
produce by itself the gap $\Delta$. For instance, the coupling
constant can be enhanced by intercalation or by some kind of a
disorder. It is pertinent to note, that FCQPT can take place in
three-dimensional metals at the usual metallic densities, as in
heavy-fermion metals, when the effective mass is sufficiently
large \cite{ps}. In that case, the potential energy can easily
prevail over the kinetic energy  leading to FCQPT at sufficiently
high densities. Then, we can expect the appearance of HTS if the
coupling constant is large to ensure high values of the
superconducting gap.

In summary, we have shown by a simple self-consistent analysis
that the general features of the shape of the critical temperature
$T_c$ as a function of the density $x$ of the mobile carriers in
metals and the value of the optimal doping $x_{opt}$ can be
understood within the framework of the theory of high-$T_c$
superconductivity based on FCQPT. We have demonstrated that
neither the $d$-wave pairing symmetry, nor the pseudogap
phenomenon, nor the presence of the Cu-O$_2$ planes are of
importance for the existence of the high-$T_c$ superconductivity.
Our theory explains the experimental observation that a strongly
correlated Fermi liquid in heavily overdoped cuprates transforms
into a normal Landau liquid. The main features of a
room-temperature superconductor have also been outlined.

This work was supported in part by the Russian Foundation for
Basic Research, No 01-02-17189.


\begin{thebibliography}{99}

\bibitem{tim} T.Timusk and B. Statt, Rep. Prog. Phys. {\bf 62}, 61
(1999).

\bibitem{tk} C.C. Tsuei and J.R. Kirtley, Rev. Mod. Phys. {\bf
72}, 969 (2000).


\bibitem{vn} C.M. Varma, Z. Nussinov, and Wim van Saarloos, Phys.
Rep. {\bf 361}, 267 (2002).

\bibitem{skin} N.-C. Yeh {\it et al.,} Phys. Rev. Lett. {\bf 87},
087003 (2001); A. Biswas {\it et al.,} Phys. Rev. Lett. {\bf 88},
207004 (2002); J.A. Skinta {\it et al.,} Phys. Rev. Lett. {\bf
88}, 207005 (2002).

\bibitem{chen} J.A. Skinta {\it et al.,} Phys. Rev. Lett. {\bf 88},
207003 (2002); C.-T. Chen {\it et al.,} Phys. Rev. Lett. {\bf 88},
227002 (2002).

\bibitem{gun} O. Gunnarsson, Rev. Mod. Phys. {\bf 69}, 575 (1997).

\bibitem{cyr} C. Proust {\it et al.,} Phys. Rev. Lett. {\bf 89},
147003 (2002).


\bibitem{and} P.W. Anderson, Science {\bf 235}, 1196 (1987).

\bibitem{kiv} S.A. Kivelson {\it et al.,} Phys. Rev. B {\bf 35},
8865 (1987).

\bibitem{lan} L. D. Landau, Zh. \'{E}ksp. Teor. Fiz. {\bf 30}, 1058
(1956) [Sov. Phys. JETP {\bf 3}, 920 (1956)].

\bibitem{ks} V.A. Khodel and V.R. Shaginyan,
Pis'ma Zh. \'{E}ksp. Teor. Fiz. {\bf 51}, 488 (1990) [JETP Lett.
{\bf 51}, 553 (1990); V.A. Khodel, V.R. Shaginyan, and V.V.
Khodel, Phys. Rep. {\bf 249}, 1 (1994).

\bibitem{vol} G. E. Volovik,
Pis'ma Zh. \'{E}ksp. Teor. Fiz. {\bf 53}, 208 (1991) [JETP Lett.
{\bf 53}, 222 (1991)].

\bibitem{ms} M.Ya. Amusia and V.R. Shaginyan,
Pis'ma Zh. \'{E}ksp. Teor. Fiz. {\bf 73}, 268 (2001) [JETP Lett.
{\bf 73}, 232 (2001)]; M.Ya. Amusia and V.R. Shaginyan, Phys. Rev.
B {\bf 63}, 224507 (2001).

\bibitem{ams} M.Ya. Amusia, S.A. Artamonov, and V.R. Shaginyan,
Pis'ma Zh. \'{E}ksp. Teor. Fiz. {\bf 74}, 396 (1998) [JETP Lett.
{\bf 74}, 435 (2001)].

\bibitem{til} D.R. Tilley and J. Tilley, {\it Superfluidity and
Superconductivity}, Bristol, Hilger (1975).

\bibitem{lpa} R.B. Laughlin and D. Pines, Proc. Natl. Acad. Sci.
U.S.A. {\bf 97}, 28 (2000); P.W. Anderson, cond-mat/007185;
cond-mat/0007287.

\bibitem{vall} T. Valla {\it et al}., Science {\bf 285}, 2110 (1999).


\bibitem{blk} P.V. Bogdanov {\it et al}., Phys. Rev. Lett. {\bf 85},
2581 (2000); A. Kaminski {\it et al}., Phys. Rev. Lett. {\bf 86},
1070 (2001).

\bibitem{val1} Z. Yusof {\it et al}., Phys. Rev. Lett. {\bf 88},
167006 (2002).

\bibitem{ms1} M.Ya. Amusia and V.R. Shaginyan,
Phys. Lett. A {\bf 298}, 193 (2002).

\bibitem{abr} A.A. Abrikosov, Phys. Rev. B {\bf 52}, R15738 (1995);
A.A. Abrikosov, cond-mat/9912394.

\bibitem{sh} V.R. Shaginyan,
Pis'ma Zh. \'{E}ksp. Teor. Fiz. {\bf 68}, 491 (1998) [JETP Lett.
{\bf 68}, 527 (1998)]; S.A. Artamonov and V.R. Shaginyan, Zh.
\'{E}ksp. Teor. Fiz. {\bf 119}, 331 (2001) [JETP {\bf 92}, 287
(2001)].

\bibitem{skb1} J.H. Sch\"on, Ch. Kloc, and B. Batlogg, Science {\bf
293}, 2432 (2001).

\bibitem{ps} Yu.G. Pogorelov and V.R. Shaginyan,
Pis'ma Zh. \'{E}ksp. Teor. Fiz. {\bf 76}, 614 (2002).

\bibitem{mac} A.P. Mackenzie {\it et al}., Phys. Rev. B {\bf 53},
5848 (1996).

\bibitem{ksz} V.A. Khodel, V.R. Shaginyan, and M.V. Zverev,
Pis'ma Zh. \'{E}ksp. Teor. Fiz. {\bf 65}, 242 (1997) [JETP Lett.
{\bf 65}, 253 (1997)].

\bibitem{mzo} N. Miyakawa {\it et al}., Phys. Rev. Lett. {\bf 83},
1018 (1999).

\end{thebibliography}
\end{document}